\documentclass[cameraready]{Interspeech}

\usepackage{amsmath,amssymb}
\usepackage{graphicx}
\usepackage{booktabs}
\usepackage{multirow}
\usepackage{array}
\usepackage{adjustbox}
\usepackage{url}

\title{Reducing Speaker Residual by Considering Pinhole Effect \\ in Voice Anonymization}

\author[affiliation={1}]{Zeyan}{Liu}
\author[affiliation={3}]{Weili}{Jiang}
\author[affiliation={1}]{Liping}{Chen}
\author[affiliation={2}]{Kong Aik}{Lee}
\author[affiliation={3}]{Boyu}{Zhao}
\author[affiliation={3}]{Kai}{Gao}
\author[affiliation={1}]{Zhenhua}{Ling}

\address{$^{1}$University of Science and Technology of China, Hefei, Anhui, China\\
$^{2}$The Hong Kong Polytechnic University, Hong Kong SAR, China\\
$^{3}$Institute of Forensic Science, Ministry of Public Security, China}

\email{xy671231@mail.ustc.edu.cn,\ jiangweili@cifs.gov.cn, \{lipchen,zhling\}@ustc.edu.cn, kong-aik.lee@polyu.edu.hk, \ zhaoboyu\_uestc@163.com,\ gaokai@cifs.gov.cn }


\makeatletter
\newcommand{\firstpagefootnote}[1]{%
  \begingroup
  \renewcommand{\thefootnote}{}%
  \footnotetext[0]{#1}%
  \endgroup
}
\makeatother

\begin{document}

\maketitle

\firstpagefootnote{%
\noindent\textit{Corresponding author: Liping Chen.}\par
\noindent\hspace*{1.5em} This work was supported in part by the National Key Research and
Development Program Project 2024YFE0217200, the Innovation and
Technology Fund of the Hong Kong SAR MHP/048/24, and the National
Natural Science Foundation of China under Grant 62506349 and U23B2053.
}

\begin{abstract}
Voice anonymization aims to protect privacy by suppressing speaker identity while preserving linguistic content and prosody. However, residual speaker attributes in non-identity representations may still increase linkability and weaken privacy protection. To this end, this paper proposes a fine-tuning strategy with a pinhole loss for well-trained voice anonymization frameworks to further reduce residual speaker attributes. Inspired by the pinhole effect, the pinhole loss measures the linkability of anonymized utterances from the same source speaker. By minimizing this loss, linkability is reduced, thereby improving privacy protection. Experiments on multiple anonymization frameworks, pseudo-speaker generation methods, and datasets show improved privacy protection while maintaining utility. Audio samples can be found in \url{https://anonymous.4open.science/r/Pinhole-loss-fine-tunning-4628}.
\end{abstract}

\section{Introduction}
Voice anonymization aims to suppress speaker identity in speech while preserving linguistic content and prosody. Its core objective is de-identification, which requires that anonymized utterances should not allow reliable speaker recognition or linkage to the original identity~\cite{vpc2024}. Most frameworks implement anonymization through representation disentanglement, where speech is disentangled into factors such as content, prosody, and speaker identity, and the identity factor is modified or replaced~\cite{vpc2024,vpc2020_result,vpc2022_result}. However, identity attributes are not perfectly separable and often leak into content and prosody representations, leaving residual speaker attributes~\cite{vpc2022_result,gaznepoglu_deep_2023}, which remain after anonymization and can still support recognition or linkage attacks. Eliminating residual leakage is difficult because speaker attributes are distributed across multiple feature streams and tightly coupled with phonetic realizations and prosodic patterns, so stronger suppression can quickly degrade intelligibility, naturalness, and prosody fidelity. Therefore, reducing residual speaker attributes is essential for reliable de-identification and remains a key technical challenge in practical voice anonymization.

Many recent studies aim to reduce residual speaker attributes by strengthening disentanglement in specific feature streams. Codec-based anonymization~\cite{nac} uses discrete quantized codes as a bottleneck to reduce residual speaker attributes, but the quantized codes may still retain substantial residual speaker information. Prosody-oriented methods modify fundamental frequency~\cite{f0_modification,meyer2023prosody,f0_mod_1,f0_mod_3} or introduce prosody cloning and control~\cite{sttts} to reduce identity leakage in prosodic patterns, yet residual cues can still persist in content related representations. Content based approaches~\cite{vq-bn} further purify acoustic model content representations using vector quantization, but they do not impose unified constraints on other streams such as prosody, where identity leakage may remain. Factorized distillation~\cite{easy} factorizes speaker identity, content, and emotion via distillation to constrain multiple factors, but residual identity can distribute across subspaces and cannot be systematically suppressed by constraining only selected components. Overall, existing methods often optimize only a subset of factors or rely on implicit bottlenecks, and they still lack a global, explicit, and optimizable objective that consistently reduces residual speaker attributes across feature streams.

In this work, residual speaker attributes are revisited from the perspective of the pinhole effect~\cite{pinhole}, which provides an intuitive view of the relationship among residual speaker attributes, linkability, and privacy protection under any-to-one pseudo-speaker mapping strategies. The key insight is that residual speaker attributes are reflected by the clustering strength of anonymized utterances from the same source speaker, and stronger clustering corresponds to higher linkability and weaker privacy protection. Motivated by this view, a fine-tuning strategy with a pinhole loss is proposed for well-trained voice anonymization frameworks. Inspired by the pinhole effect, the pinhole loss measures the compactness of anonymized utterances from the same source speaker, which reflects linkability caused by residual speaker attributes. By minimizing this loss during fine-tuning, linkability is reduced, and privacy protection is improved. Standard generation objectives are jointly retained during fine-tuning to preserve synthesis quality and utility. Experiments on multiple anonymization frameworks, pseudo-speaker generation methods, and datasets demonstrate consistent privacy gains with competitive utility preservation.

\begin{figure*}[t]
  \centering
  \resizebox{\textwidth}{!}{%
    \includegraphics{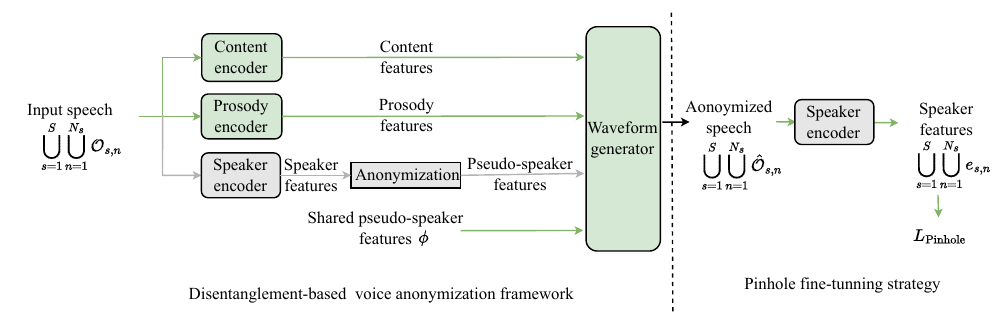}%
  }
  \caption{Overview of the voice anonymization framework and the proposed fine-tuning strategy. Modules on the left of the dashed line represent the well-trained anonymization framework, while modules on the right correspond to the Pinhole-based fine-tuning strategy. Green arrow lines are applicable in the pinhole fine-tuning, while green modules are trainable in the fine-tuning process. Gray arrow lines are inapplicable during the fine-tuning. Gray modules are frozen during fine-tuning.}
  \label{fig:overview_framework}
\end{figure*}

\section{Background}

\subsection{Overview of voice anonymization framework}
\label{sec:framework}
Figure~\ref{fig:overview_framework} shows the well-trained voice anonymization framework. Let $\mathcal{O}_{s,n}$ denote the $n$-th utterance of source speaker $s$, where the input set contains $S$ source speakers and $N_s$ utterances for speaker $s$. Given an input speech $\mathcal{O}_{s,n}$, a disentanglement-based anonymization model disentangles it into content, prosody, and speaker features using a content encoder, a prosody encoder, and a speaker encoder. The speaker features are then transformed into pseudo-speaker features by the anonymization method. Finally, a waveform generator synthesizes anonymized speech by combining the content, prosody, and pseudo-speaker features, yielding anonymized speech $\hat{\mathcal{O}}_{s,n}$.

\subsection{Revisit of pinhole effect in voice anonymization}
\label{sec:pinhole_theory}

\begin{figure}[t]
\centering
\includegraphics[width=1.0\columnwidth]{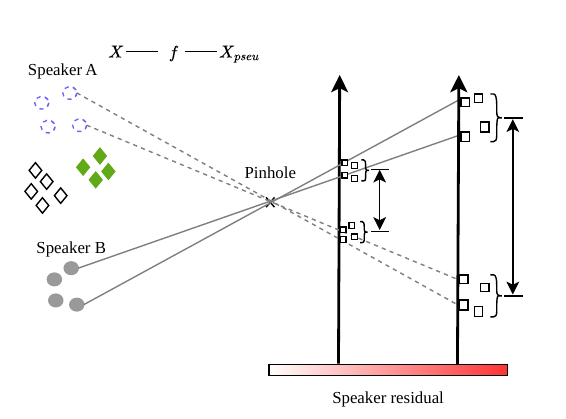}
\caption{Theoretical illustration of the pinhole effect, used as the motivation of the proposed fine-tuning strategy. In the residual bar, deeper red indicates more residual.}
\label{fig:pinhole_effect}
\end{figure}

Fig.~\ref{fig:pinhole_effect} provides an intuitive view of how residual speaker attributes affect linkability in the framework of Sec.~\ref{sec:framework}. The reader is referred to~\cite{pinhole} for the detailed theoretical derivation. The anonymization process can be viewed as a function $f$ that maps the original speaker $X$ to a pseudo-speaker $X_{\mathrm{pseu}}$. Under an any-to-one mapping, all speakers are projected toward a shared pseudo-speaker, denoted as the pinhole, into which the source speaker attributes are ideally collapsed.

In practice, imperfect disentanglement leaves residual speaker attributes in the content and prosody features, so the anonymized speech can still be clustered by source identity. As shown in Fig.~\ref{fig:pinhole_effect}, smaller residuals yield lower linkability, whereas larger residuals yield higher linkability. Motivated by this, a pinhole-based fine-tuning strategy is proposed to suppress these residuals and improve privacy protection.

\section{Pinhole-based fine-tuning}
\label{sec:proposed_ft}

In this section, the proposed pinhole-based fine-tuning strategy is presented. Different from the original voice anonymization framework used for inference, an additional fine-tuning stage is introduced to further suppress residual speaker attributes. The anonymization framework itself is kept unchanged, while a pinhole loss is introduced during fine-tuning.

\subsection{Pinhole loss}
\label{sec:pinhole_loss_definition}

As illustrated in Fig.~\ref{fig:overview_framework}, the proposed fine-tuning process is conducted on a well-trained voice anonymization framework. Let $\bigcup_{s=1}^{S}\bigcup_{n=1}^{N_s}\mathcal{O}_{s,n}$ denote the input speech set used for fine-tuning. Each input utterance is processed as described in Sec.~\ref{sec:framework}. During fine-tuning, a shared pseudo-speaker feature is used for all input speech so that residual speaker attributes in anonymized speech can be more directly reflected. The speaker encoder is applied to the anonymized speech, and speaker features $\bigcup_{s=1}^{S}\bigcup_{n=1}^{N_s} e_{s,n}$ are obtained. Based on these features, the pinhole loss is used as a measure of linkability and is computed as follows.

The global mean $\mu$ and the per-source-speaker mean $\mu_s$, where $\mu_s$ denotes the mean of the anonymized speaker embeddings sharing the same source speaker $s$, are computed as
\begin{equation}
\mu = \frac{1}{\sum_{s=1}^{S} N_s}\sum_{s=1}^{S}\sum_{n=1}^{N_s} e_{s,n},\qquad
\mu_s = \frac{1}{N_s}\sum_{n=1}^{N_s} e_{s,n}.
\label{eq:means}
\end{equation}
The within-speaker scatter $R_{\mathrm{w}}$ and between-speaker scatter $R_{\mathrm{b}}$ are defined as
\begin{equation}
R_{\mathrm{w}} = \sum_{s=1}^{S}\sum_{n=1}^{N_s}(e_{s,n}-\mu_s)(e_{s,n}-\mu_s)^{\top},
\label{eq:sw}
\end{equation}
\begin{equation}
R_{\mathrm{b}} = \sum_{s=1}^{S} N_s(\mu_s-\mu)(\mu_s-\mu)^{\top}.
\label{eq:sb}
\end{equation}
The pinhole loss is then defined as
\begin{equation}
L_{\mathrm{Pinhole}} = \frac{\mathrm{tr}(W^{\top} R_{\mathrm{b}} W)}{\mathrm{tr}(W^{\top} R_{\mathrm{w}} W)},
\label{eq:lda_ratio}
\end{equation}
where $W$ is formed by the top $k$ generalized eigenvectors of
\begin{equation}
R_{\mathrm{b}}\, w = \lambda\, R_{\mathrm{w}}\, w.
\label{eq:gev}
\end{equation}

The above loss measures the relative separability of anonymized speech across source speakers. A larger value corresponds to higher linkability, whereas a smaller value corresponds to lower linkability. Therefore, minimizing $L_{\mathrm{Pinhole}}$ reduces the separability across source speakers and suppresses residual speaker attributes.

\subsection{Fine-tuning strategy with pinhole loss}
\label{sec:pinhole_ft_strategy}

A fine-tuning stage is introduced on top of the well-trained voice anonymization framework. During fine-tuning, only the content encoder, prosody encoder, and waveform generator are updated, while the speaker encoder used for loss computation is kept frozen. This design is motivated by the main pathway of residual speaker leakage in disentanglement-based anonymization: speaker-related information may remain in content and prosody representations due to imperfect disentanglement and can be further expressed in the synthesized waveform by the waveform generator. Therefore, these three modules are fine-tuned to suppress residual speaker attributes while preserving the original anonymization mapping and pseudo-speaker assignment behavior for stable optimization. The pinhole loss is jointly optimized with the original training objectives during fine-tuning.

\section{Experiments}
\label{sec.4}

\begin{table*}[t]
    \centering
    \caption{Detailed performances of all configurations. The rows represent configurations and columns represent metrics. EERs (\%) are split by LibriSpeech subset (dev/test) and gender (f/m), with average values. WERs (\%) and UARs (\%) are presented for the corresponding evaluation datasets. All configurations compare \emph{w/o} (without) and \emph{w/} (with) the proposed fine-tuning strategy. For EER, higher is better, and the better value in each pair is bolded.}
    \label{tab:comprehensive_detailed_results}
    \setlength{\tabcolsep}{2.8pt}
    \resizebox{\linewidth}{!}{
        \begin{tabular}{l|c|ccccc|cc|cc}
            \hline
            \multirow{2}{*}{Model / Method} & \multirow{2}{*}{Pinhole-based fine-tuning} & \multicolumn{5}{c|}{EER (\%)} & \multicolumn{2}{c|}{WER (\%)} & \multicolumn{2}{c}{UAR (\%)} \\
            \cline{3-11}
            & & libri-dev-f & libri-dev-m & libri-test-f & libri-test-m & avg & libri-dev & libri-test & IEMOCAP-dev & IEMOCAP-test \\
            \hline\hline

            \multicolumn{10}{l}{\textit{x-vector based}} \\
            \hline
            a2o & \emph{w/o}  & 6.821 & 4.887 & 5.874 & 5.267 & 5.712 & 2.677 & 2.576 & 42.471 & 42.874 \\
            a2o & \emph{w/} & \textbf{21.014} & \textbf{22.607} & \textbf{21.534} & \textbf{21.445} & \textbf{21.650} & 2.641 & 2.581 & 42.621 & 42.604 \\
            \hline
            RS  & \emph{w/o}  & 12.145 & 13.679 & 13.508 & 13.974 & 13.326 & 2.695 & 2.562 & 42.578 & 42.737 \\
            RS  & \emph{w/} & \textbf{26.217} & \textbf{26.383} & \textbf{27.223} & \textbf{28.847} & \textbf{27.167} & 2.623 & 2.574 & 42.672 & 42.149 \\
            \hline
            GAN & \emph{w/o}  & 14.683 & 15.723 & 13.473 & 15.777 & 14.914 & 2.639 & 2.566 & 42.275 & 42.421 \\
            GAN & \emph{w/} & \textbf{30.401} & \textbf{28.924} & \textbf{28.741} & \textbf{29.811} & \textbf{29.486} & 2.651 & 2.608 & 42.607 & 42.551 \\
            \hline
            IDMap-Diff & \emph{w/o}  & 17.749 & 18.753 & 17.951 & 18.852 & 18.326 & 2.672 & 2.576 & 42.877 & 42.087 \\
            IDMap-Diff & \emph{w/} & \textbf{32.482} & \textbf{31.041} & \textbf{32.814} & \textbf{30.194} & \textbf{31.632} & 2.643 & 2.562 & 42.487 & 42.876 \\
            \hline\hline

            \multicolumn{10}{l}{\textit{ASRBN}} \\
            \hline
            a2o & \emph{w/o}  & 30.847 & 31.241 & 29.874 & 31.225 & 30.796 & 3.281 & 3.198 & 45.877 & 46.573 \\
            a2o & \emph{w/} & \textbf{44.574} & \textbf{46.217} & \textbf{45.384} & \textbf{45.776} & \textbf{45.487} & 3.147 & 3.301 & 46.212 & 46.941 \\
            \hline
            RS  & \emph{w/o}  & 32.647 & 33.734 & 33.455 & 33.661 & 33.374 & 3.359 & 3.266 & 45.877 & 46.573 \\
            RS  & \emph{w/} & \textbf{42.177} & \textbf{43.837} & \textbf{43.284} & \textbf{44.176} & \textbf{43.368} & 3.332 & 3.347 & 46.212 & 46.941 \\
            \hline
            GAN & \emph{w/o}  & 36.841 & 37.273 & 35.554 & 37.744 & 36.853 & 3.359 & 3.266 & 46.275 & 46.812 \\
            GAN & \emph{w/} & \textbf{45.281} & \textbf{44.871} & \textbf{42.657} & \textbf{45.868} & \textbf{44.669} & 3.251 & 3.298 & 46.447 & 46.441 \\
            \hline
            IDMap-Diff & \emph{w/o}  & 43.147 & 44.767 & 44.438 & 45.715 & 44.516 & 3.267 & 3.345 & 46.744 & 46.121 \\
            IDMap-Diff & \emph{w/} & \textbf{47.578} & \textbf{48.431} & \textbf{48.071} & \textbf{48.921} & \textbf{48.250} & 3.241 & 3.333 & 46.515 & 45.867 \\
            \hline\hline

            \multicolumn{10}{l}{\textit{ASRBN-GST}} \\
            \hline
            a2o & \emph{w/o}  & 38.747 & 40.231 & 39.543 & 39.882 & 39.600 & 3.215 & 3.12  & 51.222 & 52.047 \\
            a2o & \emph{w/} & \textbf{45.287} & \textbf{47.118} & \textbf{46.476} & \textbf{46.229} & \textbf{46.345} & 3.311 & 3.207 & 52.473 & 53.001 \\
            \hline
            RS  & \emph{w/o}  & 40.747 & 42.571 & 43.661 & 41.556 & 42.133 & 3.321 & 3.244 & 52.414 & 53.323 \\
            RS  & \emph{w/} & \textbf{45.786} & \textbf{46.711} & \textbf{46.714} & \textbf{47.221} & \textbf{46.608} & 3.279 & 3.321 & 51.861 & 53.411 \\
            \hline
            GAN & \emph{w/o}  & 45.231 & 43.672 & 44.823 & 43.248 & 44.244 & 3.377 & 3.210 & 52.554 & 51.786 \\
            GAN & \emph{w/} & \textbf{47.322} & \textbf{45.557} & \textbf{46.623} & \textbf{45.337} & \textbf{46.210} & 3.287 & 3.301 & 52.127 & 52.471 \\
            \hline
            IDMap-Diff & \emph{w/o}  & 47.507 & 48.841 & 48.602 & 47.859 & 48.202 & 3.328 & 3.306 & 52.911 & 52.988 \\
            IDMap-Diff & \emph{w/} & \textbf{49.747} & \textbf{51.071} & \textbf{52.225} & \textbf{49.974} & \textbf{50.754} & 3.277 & 3.322 & 52.817 & 53.471 \\
            \hline
        \end{tabular}
    }
\end{table*}

\subsection{Evaluation metrics}
Our evaluations were carried out following the recipes provided in VPC2024~\cite{vpc2024}\footnote{\label{fn:vpc2024-repo}\url{https://github.com/Voice-Privacy-Challenge/Voice-Privacy-Challenge-2024}}. The voice protection capability was assessed through automatic speaker verification (ASV) tests, measured by the equal error rates (EERs). The linguistic content preservation capability was examined in automatic speech recognition (ASR) tests, measured by word error rates (WERs). The capability of preserving the original emotion was evaluated with speech emotion recognition (SER), measured with unweighted average recalls (UARs). Utterances for evaluation and ASV model training were anonymized at the utterance level. For pseudo-speaker generation strategies other than \emph{a2o}, a unique pseudo-speaker was generated for every single utterance, while under \emph{a2o} the same shared pseudo-speaker was used for all utterances.

\subsection{Compared methods}

The proposed fine-tuning strategy was evaluated on two official anonymization frameworks from the VoicePrivacy Challenge 2024 (VPC2024)~\cite{vpc2024} and one open-source baseline~\cite{idmap}. These anonymization frameworks differed in how residual speaker attributes were suppressed in the content and prosody pathways.

\noindent \romannumeral 1.\ \emph{x-vector based}~\cite{average}: An x-vector based anonymization framework, where speech was factorized into content, $F_0$, and an x-vector speaker embedding, without any other method for removing residual speaker attributes from the content or prosody pathways.

\noindent \romannumeral 2.\ \emph{ASRBN}~\cite{vq-bn}: An ASR bottleneck-based anonymization framework, where speech was factorized into content, $F_0$, and an x-vector speaker embedding. Compared with the x-vector baseline, this framework further strengthened disentanglement on the content features by using ASR bottleneck features as content representations and applying vector quantization (VQ) after the bottleneck to suppressed residual speaker attributes in content.

\noindent \romannumeral 3.\ \emph{ASRBN-GST}~\cite{idmap}: Built upon \emph{ASRBN}, this framework retains VQ on the content features to remove residual speaker attributes from content representations, and further reduced speaker leakage through prosody by replacing $F_0$ with learned prosody features.

On top of each baseline, we evaluated several pseudo-speaker generation strategies to assess their effectiveness and generalization: Any to one (a2o)~\cite{pinhole}, which mapped all source speakers to the same pseudo-speaker; Random selection (RS), which followed~\cite{average} and the proximity was random; GAN-based~\cite{GAN} (GAN); Identity mapping with Diffusion-based (IDMap-Diff)~\cite{idmap}.

\subsection{Datasets and configurations}
The LibriTTS train-clean-100, train-clean-360, and train-other-500 sets~\cite{libritts} were used for training and fine-tuning for all anonymization frameworks. ASV and ASR evaluations were performed on the development and test subsets of LibriSpeech~\cite{panayotov2015librispeech}. SER evaluations were conducted on subsets derived from IEMOCAP~\cite{busso2008iemocap} following the VPC2024 configuration~\cite{vpc2024}. The pretrained ECAPA-TDNN model~\cite{ECAPATDNN}\footnote{\label{fn:ecapa-tdnn}\url{https://github.com/TaoRuijie/ECAPA-TDNN}} trained on VoxCeleb2 was used as the speaker encoder for ASRBN and ASRBN-GST anonymization frameworks. Also, it was used in fine-tuning. During fine-tuning, each batch was constructed by selecting $S=6$, and $N_s=6$, for a total of 36 utterances per batch, which were randomly sampled for each selected speaker. For pseudo-speaker generation, the any-to-one pseudo-speaker vector was set to the average of speaker vectors computed over all speakers in LibriTTS train-clean-100. For random selection, the candidate pseudo-speaker pool was formed from LibriTTS train-clean-100. The GAN and IDMap-Diff models were trained using LibriTTS train-clean-100.

\subsubsection{Voice anonymization evaluations}
The EERs, WERs, and UARs obtained in the ASV, ASR, and SER evaluations are presented in Table~\ref{tab:comprehensive_detailed_results}. The ASV evaluations were conducted in a gender-dependent manner. The average EER across the four test subsets for each method is included. Across all tested anonymization frameworks and pseudo-speaker generation strategies, the proposed fine-tuning strategy consistently improves privacy protection, as reflected by higher EERs in the paired \emph{w/o} and \emph{w/} comparisons in Table~\ref{tab:comprehensive_detailed_results}. This indicates that residual speaker attributes are broadly present across anonymization pipelines and can be effectively reduced by the proposed objective. Meanwhile, utility is largely preserved. WER changes on LibriSpeech dev and test are generally small after fine-tuning, indicating limited impact on linguistic intelligibility. UAR changes on IEMOCAP are also small overall, suggesting that emotion-related information remains comparable before and after fine-tuning. These results show that directly optimizing linkability through the pinhole loss improves de-identification without introducing large utility regressions.

\subsection{Speaker leakage probing experiment}
\label{sec:speaker_leakage_probe}

\begin{table}[t]
\centering
\caption{Speaker leakage probing (speaker classification accuracy \%) on LibriTTS dev and test. $\Delta$ is After$-$Before. Mel spectrum and $F_0$ were additionally tested as reference: Mel 94.3\%, $F_0$ 78.7\%.}
\label{tab:speaker_leakage_delta}
\scriptsize
\setlength{\tabcolsep}{4pt}
\renewcommand{\arraystretch}{1.05}
\begin{tabular}{l ccc ccc}
\toprule
\multirow{2}{*}{Baseline}
& \multicolumn{3}{c}{Content}
& \multicolumn{3}{c}{Prosody} \\
\cmidrule(lr){2-4}\cmidrule(lr){5-7}
& Before & After & $\Delta$ & Before & After & $\Delta$ \\
\midrule
x-vector based & 84.7 & 44.4 & -40.3 & \multicolumn{3}{c}{N/A ($F_0$)} \\
ASRBN  & 16.5 & 3.6  & -12.9 & \multicolumn{3}{c}{N/A ($F_0$)} \\
ASRBN-GST   & 16.7 & 3.4  & -13.3 & 36.9 & 9.3 & -37.6 \\
\bottomrule
\end{tabular}
\end{table}

ASV evaluates privacy on the final anonymized speech, but it does not reveal where speaker attributes remain within the model. Therefore, a feature-based leakage probe is used to measure residual speaker attributes in intermediate streams. Specifically, the ECAPA-TDNN\textsuperscript{\ref{fn:ecapa-tdnn}} is trained to predict the source speaker identity from a single feature stream, and the resulting classification accuracy is reported as the leakage metric, where a higher value indicates more residual speaker attributes. The probe is trained on LibriTTS train-clean-100, train-clean-360, and train-other-500, and evaluated on LibriTTS dev and test. For each baseline, content and prosody features are extracted with identical settings before and after fine-tuning. Mel spectrograms and $F_0$ are extracted from source speech and reported once as references, since they are unaffected by fine-tuning.

Table~\ref{tab:speaker_leakage_delta} reports results on LibriTTS dev and test. High reference accuracies are obtained from Mel (94.3\%) and $F_0$ (78.7\%), indicating strong speaker attribute retention. After fine-tuning, content leakage is reduced for all frameworks. For \emph{x-vector based} and \emph{ASRBN}, prosody is represented by $F_0$ that is extracted directly from the source signal and is therefore not affected by fine-tuning, which is denoted as N/A in Table~\ref{tab:speaker_leakage_delta}. For \emph{ASRBN-GST}, where prosody is encoded by a learned module, prosody leakage is strongly reduced from 36.9\% to 9.3\%. Overall, the probe confirms that residual speaker leakage in content and learned prosody representations is effectively weakened after fine-tuning, consistent with the privacy gains in Table~\ref{tab:comprehensive_detailed_results}.

\section{Conclusions}
\label{sec:conclusion}

In this work, residual speaker attributes were revisited as a key obstacle to reliable de-identification in voice anonymization. A pinhole-effect-motivated perspective was adopted. Based on this perspective, a pinhole loss was introduced to weaken linkability and used to fine-tune a well-trained voice anonymization framework. Evaluations across multiple anonymization frameworks, pseudo-speaker generation methods, and datasets showed consistently improved privacy protection, while utility was largely preserved with only small changes. These results suggest that directly optimizing linkability is an effective and practical fine-tuning strategy for reducing residual speaker attributes in existing voice anonymization pipelines.

\bibliographystyle{IEEEtran}
\bibliography{mybib}

\end{document}